\documentclass[aps,prb,showpacs,superscriptaddress,twocolumn]{revtex4-1}

\usepackage{graphicx, epsfig, amssymb, amsmath,lineno, subfigure, float,lmodern}

\begin{document}
\pagenumbering{roman}
\title{Thickness and temperature dependence of the magnetodynamic damping of pulsed laser deposited $\text{La}_{0.7}\text{Sr}_{0.3}\text{MnO}_3$ on (111)-oriented SrTi$\text{O}_3$ }

\author{Vegard Flovik}
\email{vflovik@gmail.com}
\affiliation{Department of Physics, NTNU, Norwegian University of Science and
 Technology, N-7491 Trondheim, Norway}

\author{Ferran Maci\`{a}}
\affiliation{Grup de Magnetisme, Dept. de F\'isica Fonamental, Universitat de Barcelona, Spain}
\affiliation{Institut de Ci\`encia de Materials de Barcelona (ICMAB-CSIC), Campus UAB, 08193 Bellaterra, Spain}

\author{Sergi Lend\'inez}
\affiliation{Grup de Magnetisme, Dept. de F\'isica Fonamental, Universitat de Barcelona, Spain}

\author{Joan Manel Hern\`{a}ndez}
\affiliation{Grup de Magnetisme, Dept. de F\'isica Fonamental, Universitat de Barcelona, Spain}

\author{Ingrid Hallsteinsen}
\affiliation{ Department of Electronics and Telecommunications, NTNU, Norwegian University of Science and Technology, N-7491 Trondheim, Norway}

\author{Thomas Tybell}
\affiliation{ Department of Electronics and Telecommunications, NTNU, Norwegian University of Science and Technology, N-7491 Trondheim, Norway}

\author{Erik Wahlstr\"{o}m}
\affiliation{Department of Physics, NTNU, Norwegian University of Science and
  Technology, N-7491 Trondheim, Norway}

\date{\today}

\begin{abstract}
We have investigated the magnetodynamic properties of $\text{La}_{0.7}\text{Sr}_{0.3}\text{MnO}_3$ (LSMO) films of thickness 10, 15 and 30 nm grown on (111)-oriented  SrTi$\text{O}_3$ (STO) substrates by pulsed laser deposition. Ferromagnetic resonance (FMR) experiments were performed in the temperature range 100--300 K, and the magnetodynamic damping parameter $\alpha$ was extracted as a function of both film thickness and temperature.   
We found that the damping is lowest for the intermediate film thickness of 15 nm with $\alpha \approx 2 \cdot 10^{-3}$, where $\alpha$ is relatively constant as a function of temperature well below the Curie temperature of the respective films.

\end{abstract}


\maketitle

\section{Introduction}
The magnetodynamic properties of nanostructures have received extensive attention, from both fundamental and applications viewpoints \cite{magnetism_review, magnetism_review2, magnetism_review3}. 
Nanometer sized magnetic elements play an important role in advanced magnetic storage schemes \cite{magn_recording,mram}, 
and their static and most importantly their dynamic magnetic properties are being intensely studied \cite{magndyn,magndyn2,magndyn3}.

Complex magnetic oxides display intriguing properties that make these materials promising candidates for spintronics and other magnetic applications \cite{oxide_intro}.
Manganites have received attention due to a large spin polarization, the appearance of colossal magneto-resistance and a Curie temperature above room temperature \cite{LSMO1,LSMO2,LSMO3,LSMO4}.
Within manganites, LSMO has been regarded as one of the prototype model systems. Transport and static magnetic properties of LSMO are well studied, but less attention has been paid to the magnetodynamic properties. 
For applications in magnetodynamic devices a low magnetic damping is desirable, and having well defined magnetic properties when confined to nanoscale dimensions is crucial.

The dynamic properties can be investigated by ferromagnetic resonance spectroscopy (FMR), which can be used to extract information about e.g. the effective magnetization, anisotropies and the magnetodynamic damping. Earlier studies on LSMO have investigated the dynamic properties of the magnetic anisotropies \cite{LSMO_anis}, also providing evidence for well defined resonance lines, which is a prerequisite for magnetodynamic devices. 

The magnetodynamic damping is an important material parameter that can be obtained through FMR spectroscopy by measuring the resonance linewidth as a function of frequency.
The linewidth of the resonance peaks has two contributions; an inhomogeneous contribution that does not depend on the frequency, and the dynamic contribution that is proportional to the precession frequency and to the damping parameter $\alpha$. 

Earlier studies by Luo \textit{et. al.} have investigated the magnetic damping in LSMO films grown on (001)-oriented STO capped by a normal metal layer \cite{LSMO_damping, LSMO_ISHE}, and found a damping parameter of $\alpha \approx 1.6 \cdot 10^{-3}$ for a 20 nm thick LSMO film at room temperature. Typical ferromagnetic metals have damping values of  $\alpha \approx 10^{-2}$, and the low damping indicate LSMO as a promising material for applications in magnetodynamic devices. 
The studies by Luo \textit{et. al.} were performed at room temperature, whereas the Curie temperature of LSMO is around 350 K. However, the Curie temperature depends strongly on film thickness and approaches room temperature as the thickness is decreased. Being able to control the temperature is thus important in order to accurately characterize the damping in thin LSMO films.

\begin{figure*}[]
\centering
\includegraphics[width=160 mm]{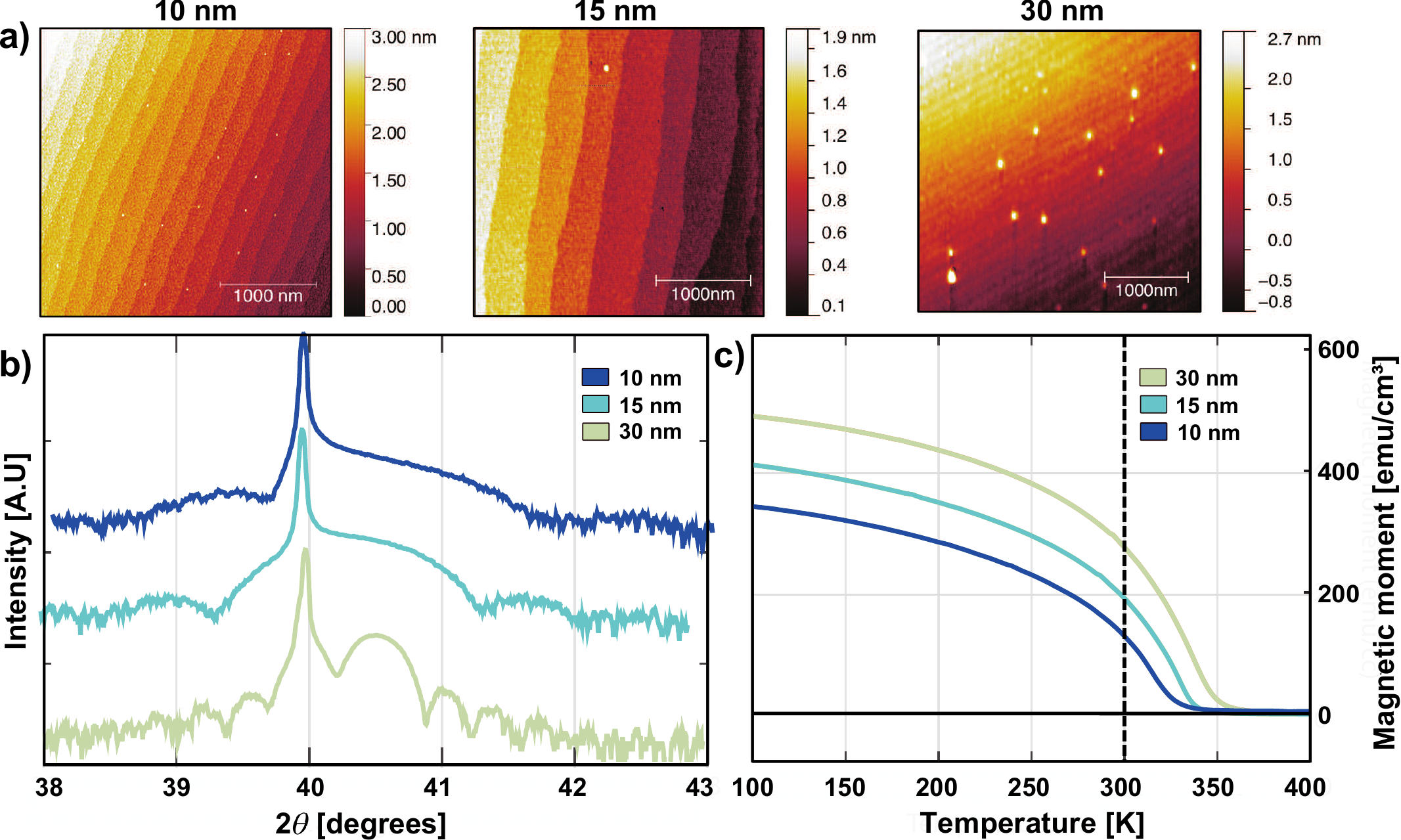}
\caption{\footnotesize a) AFM topography for films of thickness 10 nm, 15 nm and 30 nm.  b) XRD $\theta - 2\theta$ scans of the Bragg reflection for the respective film thicknesses. c) Magnetic moment measurements (performed with a vibrating sample magnetometer, (VSM)), showing the difference in magnetic moment and Curie temperature for the various film thicknesses.   }
\label{fig:sample_char}
\end{figure*}

The thickness and temperature dependence of static and dynamic magnetic properties of thin film LSMO grown on (001)-oriented  STO have been investigated in a previous study by Monsen \textit{et. al} \cite{LSMO5}. The dynamic properties were characterized from the FMR linewidth measured in a cavity based FMR setup at a fixed frequency of 9.4 GHz, and provided evidence that the magnetic damping is dominated by extrinsic effects for thin films. 

The properties of complex magnetic oxides are very sensitive to the structural parameters, hence thin film growth can be used to engineer the magnetic properties. We have previously shown that LSMO grown on (001)-oriented STO results in a biaxial crystalline anisotropy, compared with almost complete in-plane isotropy for LSMO grown on (111)-oriented STO \cite{LSMO_6}. 
The possibility to control the functional properties at the nm-scale make these materials promising for spintronics and other magnetic based applications. Hence, detailed studies on how film thickness affect the magnetic properties is important.

Here, we investigate the magnetodynamic properties of pulsed laser deposited LSMO films of thickness 10, 15 and 30 nm grown on (111)-oriented  STO (in contrast to the previous study by Monsen \textit{el. al} for LSMO grown on  (001)-oriented  STO \cite{LSMO5}).
By performing broadband FMR experiments we separate the inhomogeneous linewidth broadening and the dynamic contribution to the FMR linewidth, and extract the magnetodynamic damping parameter $\alpha$.
The main objective of our study is to characterize $\alpha$ for the various film thicknesses. 
However, as $T_c$ changes with film thickness, it is difficult to compare the absolute values at a single temperature.
We thus performed experiments in the temperature range T=100--300 K using the FMR setup described in section \ref{sec:exp_setup}, allowing us to extract $\alpha$ as a function of both temperature and film thickness.

\section{Sample growth and experimental setup}
\subsection{Sample growth and characterization}

 $\text{La}_{0.7}\text{Sr}_{0.3}\text{MnO}_3$ thin films were deposited by pulsed laser deposition on (111)-oriented SrTi$\text{O}_3$ substrates. 
A KrF excimer laser (=248 nm) with a fluency of $\approx$ 2 Jcm$^{-2}$ and repetition rate 1 Hz was employed, impinging on a stoichiometric $\text{La}_{0.7}\text{Sr}_{0.3}\text{MnO}_3$ target. The deposition took place in a 0.35 mbar oxygen ambient, at 500 Celsius and the substrate-to-target separation was 45 mm, resulting in thermalized ad-atoms \cite{LSMO_sampleprep, LSMO_sampleprep2}. After the deposition, the films were cooled to room temperature in 100 mbar of oxygen at a rate of 15 K/min. Atomic force microscopy (AFM) was used to study the surface topography. The AFM topography images  shown in Fig. \ref{fig:sample_char}a confirm the step and terrace morphology of the films after growth for the 10 nm and 15 nm thick films. For the 30 nm film, we observe a surface relaxation and transition to a more 3 dimensional growth and rougher surface compared to the 2 dimensional layer by layer growth for thinner films. 

The crystalline structure of the films was investigated using x-ray diffraction (XRD), and the XRD scans of the respective films are shown in Fig. \ref{fig:sample_char}b.

Magnetic moment measurements were performed with a vibrating sample magnetometer (VSM). In Fig. \ref{fig:sample_char}c the temperature dependence of the saturation magnetization is plotted against temperature, taken during warm-up after field cooling in 2 T. Both saturated moment and $T_c$ increase with film thickness as expected for thin films, and the values are comparable to similar thicknesses of LSMO films grown on (001)-oriented SrTi$\text{O}_3$ \cite{LSMO5}.

\subsection{FMR experiments}\label{sec:exp_setup}

The FMR experiments were performed using a vector network analyzer (VNA) setup in combination with a coplanar
waveguide (CPW) that created microwave magnetic fields of different frequencies to the film's structure. 
We used a He cryostat with a superconducting magnet capable of producing bipolar bias fields up to 5 T. The experiments were performed in the temperature range T=100--300 K using a CPW designed specially for the cryostat and using semi-rigid coaxial cables capable to carry up to 50 GHz electrical signals.

The static external field, $H_0$, was applied in the sample plane, and perpendicular to the microwave fields from the CPW. 
We measured the microwave transmission and reflection parameters as a function of field and frequency in order to obtain a complete map of the ferromagnetic resonances. Magnetic fields were varied from -500 to 500 mT, and microwave frequencies from 1 to 20 GHz.

By measuring the microwave absorption as a function of both microwave frequency and the applied external field, one can obtain the FMR dispersion. The field vs. frequency dispersion when the field is applied in the film plane is described by the Kittel equation \cite{Kittel} given by the following: 

\begin{equation}\label{eq:Kittel}
f_{\text{FMR}}=\frac{\gamma}{2 \pi} \sqrt{H_0(H_0 + H_{\text{eff}})}.
\end{equation}
Here, $f_{\text{FMR}}$ is the FMR frequency and $\gamma$ is the gyromagnetic ratio, where $\gamma / 2\pi \approx$ 28 GHz/T. $H_0$ is the applied external field and $H_{\text{eff}}=4\pi M_s -H_k $ is the effective field given by the saturation magnetization $M_s$ and the anisotropy field $H_k$.

The FMR absorption lineshape is often assumed to have a symmetric Lorentzian lineshape. However, in conductive samples, induced microwave eddy currents in the film can affect the linshape symmetry \cite{lineshape1,lineshape2}. In an experimental setup containing waveguides, coaxial cables etc., the relative phase between the electric and magnetic field components can also affect the lineshape \cite{lineshape3, lineshape4}.
We thus fit the FMR absorption, $\chi$, to a linear combination of symmetric and antisymmetric contributions, determined by the $\beta$ parameter in Eq.\ (\ref{eq:fiteq}). 

\begin{equation}\label{eq:fiteq}
\chi = A \frac{1+ \beta (H_R-H_0)/\Delta H}{(H_R-H_0)^2 + (\Delta H /2)^2}.
\end{equation}
\noindent

Here A is an amplitude prefactor, $H_R$ and $H_0$ are the resonance field and external field respectively and $\Delta H$ the full linewidth at half maximum  (FWHM). 
By measuring the FMR linewidth as a function of the microwave frequency $f_{\text{mw}}$, we extract the damping parameter $\alpha$  and the inhomogeneous linewidth broadening $\Delta H_0$ from the following relation \cite{linewidth}:

\begin{equation}\label{eq:damping}
\Delta H=   \frac{4 \pi}{\gamma}   \alpha  f_{\text{mw}} + \Delta H_0. 
\end{equation}

\section{Results and discussion}

\begin{figure*}[]
\centering
\includegraphics[width=180 mm]{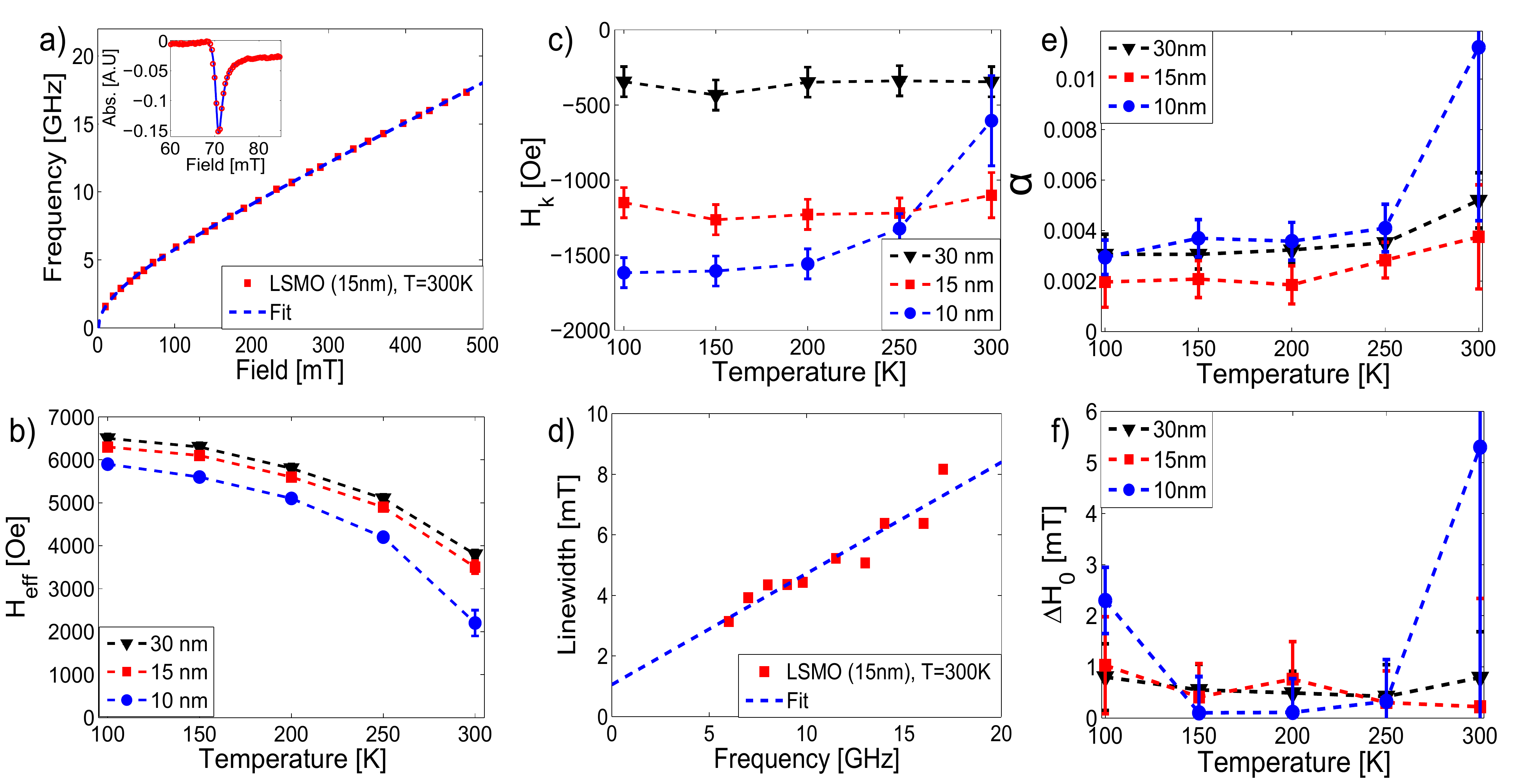}
\caption{\footnotesize a) FMR frequency vs. external magnetic field. Experimental datapoints as red squares, and fit to  Eq.\ (\ref{eq:Kittel}) as dotted line. Inset: Typical absorption lineshape and the Fit to Eq.\ (\ref{eq:fiteq}). b) Extracted values of $H_{\text{eff}}$ from the fit to Eq.\ (\ref{eq:Kittel}). c) Calculated anisotropy fields $H_k$. d) FMR absorption linewidth vs. frequency.  Experimental datapoints shown as red squares, and the fit to Eq.\ (\ref{eq:damping}) as dotted line.  e) Extracted damping parameter $\alpha$ and f) inhomogeneous linewidth broadening $\Delta H_0$ as a function of film thickness and temperature.  }
\label{fig:damping}
\end{figure*}

The measured FMR absorption shows a good agreement with the Kittel dispersion given by Eq.\ (\ref{eq:Kittel}).
In Fig. \ref{fig:damping}a we show as an example the FMR dispersion for the 15 nm film measured at room temperature, with the fit to
Eq.\ (\ref{eq:Kittel}) as dotted line. A typical absorption lineshape and the fit to Eq.\ (\ref{eq:fiteq}) is shown as inset.
Fitting the FMR dispersion to Eq.\ (\ref{eq:Kittel}) allows us to extract the effective field $H_{\text{eff}}$, given by the saturation magnetization $M_s$ and the anisotropy field $H_k$ through the relation $H_{\text{eff}}=4\pi M_s -H_k $. The extracted values of $H_{\text{eff}}$ for the various films are shown in Fig. \ref{fig:damping}b. From $H_{\text{eff}}$ and $M_s$ (shown in Fig. \ref{fig:sample_char}c) we calculate the anisotropy field $H_{k}$. The obtained values for $H_k$ are shown in Fig. \ref{fig:damping}c and indicate a strain induced negative perpendicular magnetocrystalline anisotropy (PMA), as expected for epitaxial films. The negative PMA is in agreement with that observed in LSMO grown on (001)-oriented STO \cite{LSMO_PMA}. As indicated in Fig.\ \ref{fig:damping}c, the PMA is strongest for the thinnest film and decreases as one approaches the Curie temperature. 

The main objective of our study is to characterize the magnetodynamic damping, as measured through the FMR linewidth. As an example we show in Fig.\ \ref{fig:damping}d the linewidth vs. frequency for the 15 nm sample at T=300 K. The linear relation between linewidth and frequency from Eq.\ (\ref{eq:damping}) allows us to extract the damping parameter $\alpha$ and the inhomogeneous linewidth broadening $\Delta H_0$.

To ensure all samples are in a fully ferromagnetic (FM) state, we first compare the damping well below $T_c$ of the respective films.
We find that the damping is relatively constant as a function of temperature well below $T_c$, as shown in Fig.\ \ref{fig:damping}e, and the lowest damping was found for the intermediate film thickness of 15 nm with $\alpha \approx 2\cdot 10^{-3}$. 
This value of $\alpha$ is comparable to that found in a previous study by Luo \textit{et. al.} \cite{LSMO_damping} for an LSMO film of thickness 20 nm, but in their case grown on (001)-oriented STO.

We observe an increased damping as the temperature approaches the Curie temperature of the respective films, which we  attribute to the coexistence of ferromagnetic and paramagnetic domains as $T \rightarrow T_c$. The increase in damping is largest for the 10 nm film, in agreement with the reduced $T_c$ for this film compared to the thicker 15 nm and 30 nm films, as shown in Fig.\ \ref{fig:sample_char}c. This behavior is consistent with data from LSMO single crystals \cite{LSMO_linewidth}, and with previous work by Monsen \textit{et. al. } \cite{LSMO5} for LSMO films grown on (001)-oriented STO.

The difference in damping for the various film thicknesses is observed well below $T_c$ (see Fig.\ \ref{fig:damping}e), and is thus not caused by the coexistence of ferromagnetic and paramagnetic domains. 
The increased damping for the 10 nm film compared to the 15 nm film is rather attributed to the increased importance of surface defects for the thinnest film. Surface/interface imperfections/scatterers induce a direct thickness dependent broadening due to local variations in resonance field, or indirectly through two-magnon processes, as these contributions become more dominating as the film thickness is reduced \cite{twomagnon}. Previous studies of the thickness dependence of the FMR linewidth in LSMO films grown on (001)-oriented STO by Monsen \textit{et. al.} \cite{LSMO5} show similar behavior, with a minimum in the linewidth for an intermediate film thickness and increased linewidth for thicknesses below approximately 10 nm. 

For the 30 nm film we also observe an increased damping compared to the 15 nm film. Interface effects should be less important for the 30 nm film, and the increased damping is attributed to other effects as the thickness is increased. It is known that LSMO films can experience strain relief relaxation as the film thickness is increased \cite{LSMO_strain,LSMO_strain2}, and for the 30 nm film we observe a rougher surface compared to the thinner films. This change in film structure can be observed in the AFM topography images in Fig. \ref{fig:sample_char}a, and the difference in film structure could thus cause an increased damping for the thickest film. 
Another consideration is the eddy-current contribution to the damping in conducting films \cite{EC_linewidth}. The eddy-current contribution scales with the film thickness $d$, as $d^2$, and separating the various contributions to the damping would thus require a more detailed study of the scaling of damping vs. film thickness.

In a homogeneous strain-free ferromagnet one expects that the inhomogeneous linewidth broadening, $\Delta H_0$, should be independent of temperature well below $T_c$ \cite{LSMO_tempdep}. 
The relatively temperature independence of $\Delta H_0$ for the 15 nm and 30 nm films shown in Fig. \ref{fig:damping}f indicate high quality samples, with $\Delta H_0 < 1$ mT. 
For the 10 nm film there is a slight increase in $\Delta H_0$ at low temperature, with an inhomogenous broadening of $\Delta H_0 \approx 2$ mT. The increase in $\Delta H_0$ at 300 K is attributed to the reduced $T_c$ of the 10 nm film compared to the 15 and 30 nm films, and the coexistence of ferromagnetic and paramagnetic domains as $T \rightarrow T_c$.

\section{Summary}

We have characterized the magnetic damping parameter $\alpha$ in 10, 15 and 30 nm thick LSMO films grown on (111)-oriented STO for temperatures $T$=100--300 K.
We found that $\alpha$ is relatively independent of temperature well below the Curie temperature of the respective films, with a significant increase as $T \rightarrow T_c$  due to the coexistence of ferromagnetic and paramagnetic domains.
The lowest damping was found for the intermediate film thickness of 15 nm, with $\alpha \approx 2 \cdot 10^{-3}$. For the 10 nm film, we attribute the increased damping to the increased importance of surface/interface imperfections/scatterers for thinner films. For the 30 nm film the increased damping is attributed to changes in the film structure with an increased surface roughness compared to the thinner films, as well as additional eddy-current contributions to the damping as the film thickness is increased. 

The damping of $\alpha \approx 2 \cdot 10^{-3}$ for the 15 nm film  is lower than that of e.g Permalloy ($\text{Ni}_{80}\text{Fe}_{20}$) which is one of the prototype materials for magnetodynamic devices. The low damping, in addition to other intriguing material properties like large spin polarization, indicate LSMO as a promising material for applications in magnetodynamic devices.

\section*{Acknowledgements}
This work was supported by the Norwegian Research Council (NFR), project number 216700. 
V.F acknowledge partial funding obtained from the Norwegian PhD Network on Nanotechnology for Microsystems, which is sponsored by the Research Council of Norway, Division for Science, under contract no. 221860/F40. 
F.M. acknowledges financial support from RYC-2014-16515 and from the
MINECO through the Severo Ochoa Program (SEV- 2015-0496). 
J.M and F.M aslo acknowledge funding from MINECO through MAT2015-69144.

\end{document}